# Comparing affective responses to standardized pictures and videos: A study report

Marko Horvat[1], Davor Kukolja[2] and Dragutin Ivanec[3]

[1]Polytechnic of Zagreb, Department of Computer Science and Information Technology
[2]University of Zagreb, Faculty of Electrical Engineering and Computing, Department of Electric Machines, Drives and Automation
[3]University of Zagreb, Faculty of Humanities and Social Sciences, Department of Psychology
E-mail: marko.horvat@tvz.hr

**Abstract** - Multimedia documents such as text, images, sounds or videos elicit emotional responses of different polarity and intensity in exposed human subjects. These stimuli are stored in affective multimedia databases. The problem of emotion processing is an important issue in Human-Computer Interaction and different interdisciplinary studies particularly those related to psychology and neuroscience. Accurate prediction of users' attention and emotion has many practical applications such as the development of affective computer interfaces, multifaceted search engines, video-on-demand, Internet communication and video games. To this regard we present results of a study with $N$=10 participants to investigate the capability of standardized affective multimedia databases in stimulation of emotion. Each participant was exposed to picture and video stimuli with previously determined semantics and emotion. During exposure participants' physiological signals were recorded and estimated for emotion in an off-line analysis. Participants reported their emotion states after each exposure session. The *a posteriori* and *a priori* emotion values were compared. The experiment showed, among other reported results, that carefully designed video sequences induce a stronger and more accurate emotional reaction than pictures. Individual participants' differences greatly influence the intensity and polarity of experienced emotion.

## I. INTRODUCTION

Any multimedia file can generate positive, negative or neutral emotions of varying intensity and duration [1]. By observing still images, films, printed text or listening to sounds, music and voices emotional states of affected subjects may be modulated [2] [3]. This spontaneous cognitive process is an important research topic in psychology, neuroscience and cognitive sciences but also in many interdisciplinary domains like Affective Computing and Human-Computer Interaction (HCI).

Multimedia documents with *a priori* annotated semantic and emotion content are stored in affective multimedia databases and are intended for inducing or stimulating emotions in exposed subjects. Because of their purpose such multimedia documents are also referred to as stimuli. Affective multimedia databases are standardized which allows them to be used in a controllable and predictable manner: the emotion elicitation results can be measured, replicated and validated by different research teams [4] [5]. Combined with immersive and unobtrusive visualization hardware in low-interference ambient affective multimedia databases provide a simple, low-cost and efficient means to investigate a wide range of emotional reactions [6] [7].

Compared to static pictures and sound, video is more powerful format for elicitation of emotional states because it can seamlessly and concurrently stimulates visual and auditory senses thereby multiplying their individual impacts through psychophysiological and neurological underlying mechanisms [8] [9]. It has been repetitively experimentally demonstrated that if careful attention is paid to video editing, i.e. the temporal and contextual alignment of multimedia stream relative to personal cognitions of targeted subjects, it is possible to achieve more intense and accurate stimulation of emotional states and related phenomena such as attention, anxiety and stress [7]. In practical terms affective video databases are much more useful tools than picture of sound databases. However, today very few affective video databases exist while the most prevalent are picture databases. Therefore, it is important to explore the possibility and scope of using documents from existing picture and sound databases to construct successful video sequences for fast, accurate and strong stimulation of emotional states. This goal was behind the motivation for the described study.

The remainder of this paper is organized as follows; Section 2 provides background information about the experimental study and its setup. Section 3 brings forward the results of the study which are discussed in Section 4. Finally, Section 5 concludes the paper and outlines future work into this subject.

## II. METHOD

The study was performed at University of Zagreb, Faculty of Electrical Engineering and Computing in cooperation with experts from Department of Psychology, Faculty of Humanities and Social Sciences. A homogeneous group of $N$=10 college students (4 males, 6 women) with an average age 23.8 years ($std$ = 4.35) participated in the experiment.

Each participant was stimulated with videos and still images taken from the International Affective Picture System (IAPS) [4] and the International Affective Digital Sounds System (IADS) [5]. IAPS and IADS are two of the most cited databases in the area of affective stimulation. These databases were created with three goals in mind: *i)* better experimental control of emotional stimuli, *ii)* increasing the ability of cross-study comparisons of results, and *iii*) facilitating direct replication of undertaken studies [10]. In this experiment a picture from IAPS and a sound from IADS were combined to make one video-clip. Same IAPS pictures were also used as still image stimuli without sounds. Some of the pictures used to construct the emotion elicitation sequences are shown in Fig. 1.

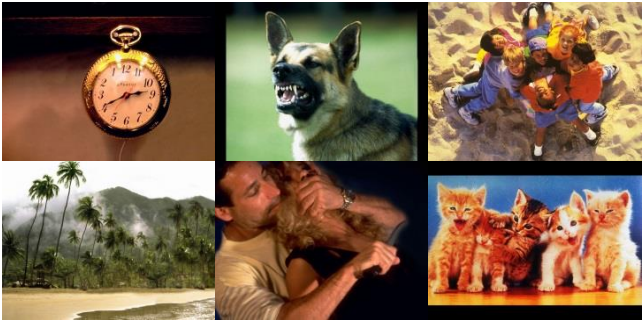

Fig. 1. A sample of IAPS pictures used as emotion eliciting video clips and images. Neutral (left column), fear (middle) and happiness dominant emotion stimuli (right).

The dominant emotions purposely elicited in the experiment were happiness and fear. The stimuli were selected using published research on emotion norms in IAPS and IADS [11] [12] as the most powerful images and sounds for simulation of the two targeted emotions. Firstly, using 200 images and 100 sounds were selected and ranked based on their intensity of happiness and fear emotion norms [11] [12]. Secondly, the sorted list was thoroughly examined by psychology experts and 20 optimal pictures and 20 sounds were manually selected for inclusion in the elicitation sequences. These stimuli were considered the most likely to induce happiness and fear in the participants' population.

Each participant was exposed to four emotion elicitation sequences in two separate sessions or series. Each session consisted of one happiness sequence and one fear inducing sequence, and also of one video sequence and one still image sequence. A single sequence was made up from 10 images or 10 video-clips (Fig. 2). Therefore, in total each participant was exposed to 20 images and 20 video-clips. The length of each stimulus was exactly 15 seconds after which the participant was shown a blank neutral screen and had to write down his affective judgments in a self-assessment questionnaire (SAQ). The expression of subjects' ratings was not time restricted after which the participant could resume the sequence by himself (i.e. with a mouse click). Immediately before the start of the experiment each participant was separately introduced to the stimulation protocol with a neutral sequence. The neutral sequence consisted of one low arousal and valence picture (N Pic) and one video-clip without dominant emotions (N Video). All stimulation sequences are available by contacting the first author.

Half of the participants were first exposed to happiness sequences, and then fear sequences, and also a half of the participants first watched still images and then videos. To prevent the unwanted drift of physiological signals (cardiac and respiratory) before nonneutral sequences participants were exposed to a neutral stimulus until their baseline response was established [13]. The neutral blank screen only showed teal color which − according to [14] − has an optimal ratio of stimulating positive and negative emotions.

The participants' emotional responses were recorded by two methods: 1) self-assessment responses i 2) real-time monitoring of physiological signals. After each exposure to a stimulus participants filled out a self-assessment questionnaire. Each report was unique to a specific stimulus and participants could not see its contents before the stimulus has finished. The test contained following instructions: 1) "Evaluate the type and intensity of emotions" for each of the emotional norms ("happiness", 'sadness", "disgust", "fear" and "some other emotion" if none of previous) on a scale with values 0 − 9, where 0 represented "None", 9 "Extremely" while 5 was a neutral value; and 2) "Evaluate pleasure and arousal" with values -4 – 4 where value -4 was labeled "Extremely unpleasant" and "Extremely calming", and 4 "Extremely pleasant" and "Extremely arousing". Value 0 indicated a neutral sensation of valence or arousal. The report was validated during the preparations for the experiment.

The monitored physiological signals were skin conductance, electrocardiogram (ECG), respiration and skin temperature with a sampling frequency of 1250 Hz. For the acquisition of signals we used BIOPAC MP150 with AcqKnowledge software. The system was synchronized with SuperLab tool for presentation of stimuli to the participants. Emotional states were estimated off-line, with varying levels of certainty, from the recorded signals using previously developed MATLAB software [15] [16]. Before starting the experiment, each participant read the instructions, filled introductory questionnaire and signed informed consent agreement. Additional help, if necessary, was provided by the trained lab assistant who also placed physiological sensors on the participant's body.

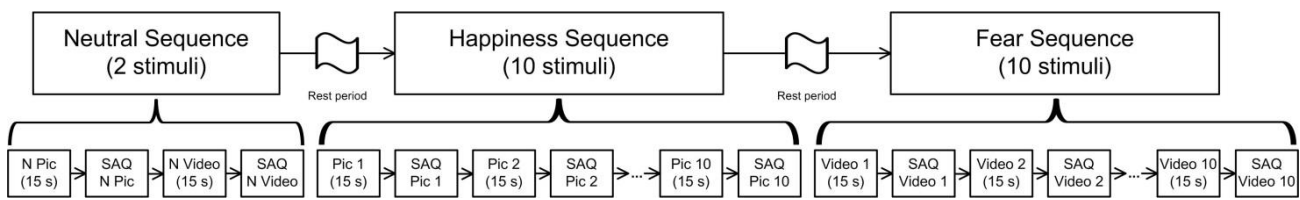

Fig. 2. The timeline of emotion stimulation paradigm. The order of dominant emotion (happiness/fear) and multimedia format (picture/video) were randomized.

Participants were seated in a separate part of the lab, 60 – 90 cm before a 19" 4:3 high-definition LCD computer monitor and wore ergonomic headphones. The supervisor station was equipped with a separate computer monitor where the experiment was controlled and participants' psychological signals were monitored in real-time [16]. The experimenter and participants were physically separated by a solid artificial wall and had no mutual visual or auditory contact. Additionally, participants did not experience sensory distractions during the experiment. The implemented procedures and the experimental layout were compatible with a setup of a common psychophysiological laboratory [17].

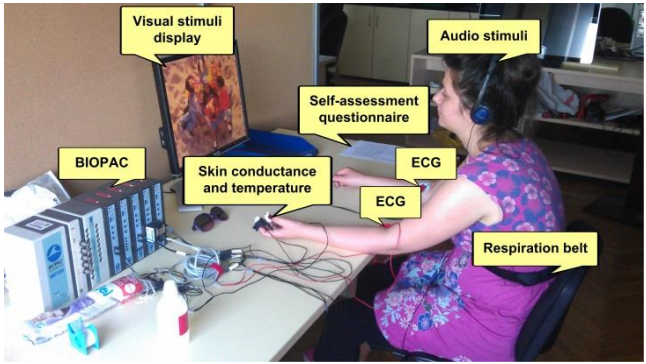

Fig. 3. The layout of participants' station. A person is observing LCD screen with audio headphones and wearing skin conductance, electrocardiogram, respiration and skin temperature contact sensors connected to BIOPAC MP150 system. Self-assessment questionnaire is on the desk. Supervisor is seated on the other side of the wall barrier.

A rest period followed each exposure session during which participants relaxed. This was verified by examining the physiological signal parameters that were visualized in real time. The exposure could resume only after the baseline signal levels were reestablished.

## III. RESULTS

The potential of evoking emotional reactions using video clips − constructed from IAPS pictures and IADS sounds − and IAPS pictures was evaluated under equal conditions. Emotional dimensions pleasure and arousal were rated on a scale 1 − 9, and emotion norms (i.e. discrete emotions) on a scale 1 − 10. A lower value (1 − 3) in both scales implies a lesser experience and a higher value (7 − 10) a more intense experience of the particular emotional dimension or norm, respectively. Also, higher frequency of ratings in the upper part of the scale (response > 5) or the highest attainable values (9 and 10) signifies a more intense and powerful stimulation. The frequencies of responses are displayed as frequency distribution diagrams. The aggregated results are shown in Fig. 4.

## IV. DISCUSSION

Results based on the analysis of participants' self-assessments indicate that the most pronounced reported difference is in arousal emotion dimension. Indeed, videos relative to pictures can more frequently stimulate higher arousal in sequences with dominant happiness and fear. Additionally, video sequences often stimulated higher levels of pleasure in happiness sequences, and lower levels of pleasure in fear sequences, but this distinction is less pronounced than with arousal.

From Fig. 4 it is evident that video sequences were more powerful in stimulation of both emotional dimensions than picture sequences. Happiness-dominant video sequences more often elicited higher levels of happiness basic emotion than happiness-dominant picture sequences. Similarly, fear-dominant video sequences provoked more above average fear ratings than fear-dominant picture sequences. Although frequency distribution differences in basic emotions are present, they are less obvious than the spread in emotional dimensions, especially arousal. In general, the emotion provoking potential of video is more apparent in emotional valence and arousal than in specific emotions happiness, sadness, disgust and fear. Due to reported low stimulation of other emotions except fear and happiness it may be concluded that the sequences were emotionally accurate. This is particularly evident in very low levels of disgust and sadness in happiness sequences and even in fear sequences. In overall video sequences provoked a "cleaner" and more powerful affective response, i.e. with lower reported intensities of emotions different from those targeted, than picture sequences which corresponds well with findings from previous studies [13] [18].

However, because of relatively small number of participants the results analysis is strongly influenced with noticeable differences between individual reports. There is a significant variability in the intensity of provoked emotion, polarity (i.e. positive or negative) and discrete category among some participants. Such discrepancies are present in video and picture sequences.

Unfortunately, due to objective reasons it was not possible to include more participants in the study. Subsequently, an independent stimulation protocol could not be implemented and the same visual stimuli had to be used in video and image sessions. If the number of participants was significantly larger different stimuli could be used in videos and images.

Based on the collected results it can be expected that multimedia sequences, carefully prepared for a particular group of participants, will be able to provoke targeted emotional states with the desired intensity. However, construction of optimal sequences proved to be difficult because IAPS, and particularly, IADS databases do not have a wide selection of stimuli with accentuated specific basic emotions. A better choice of provoking visual stimuli is clearly needed which encourages construction of more affective multimedia databases, annotated both with emotional dimensions and discrete emotions, and having a large semantic space. This in turn necessitates development of powerful tools for multimedia stimuli retrieval which can efficiently perform multifaceted search in such databases, along several emotional, semantic and contextual data dimensions, thus assisting researchers in finding optimal stimuli for personalized emotion elicitation sequences [19].

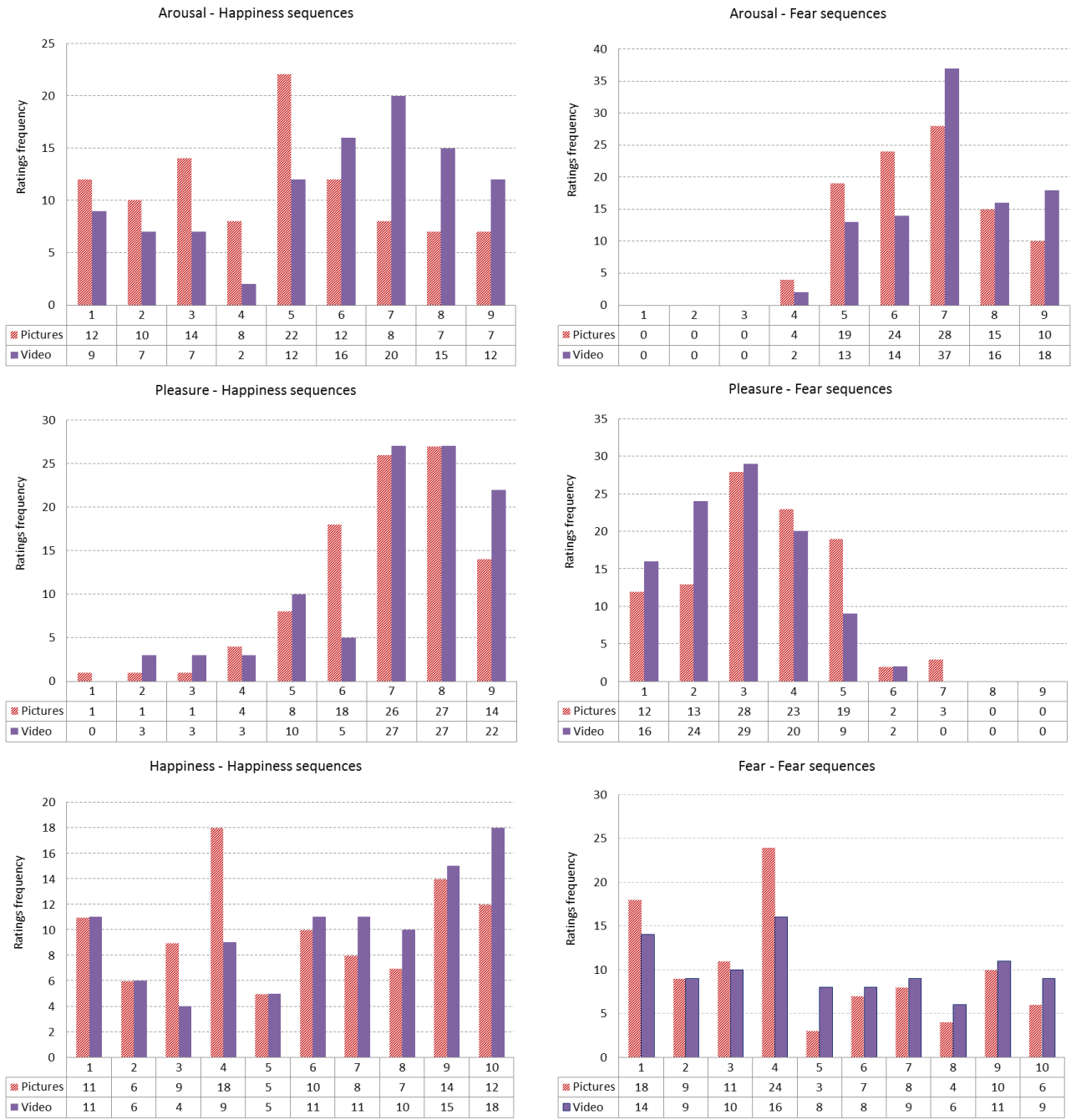

Fig. 4. Frequency distribution of *N*=10 participants' ratings after elicitation with IAPS and IADS video clips ("Video") and IAPS pictures ("Pictures"). Reported emotional dimensions arousal and pleasure (upper and middle rows), discrete emotions happiness and fear (bottom row).

## V. CONCLUSION AND FUTURE WORK

Emotional reactions can be induced by virtually any multimedia format: films, pictures, sounds, voice and even text. Participants' responses depend not only on stimuli content, properties and type, but also on a number of intrinsic and extrinsic factors which may be difficult to control in a practical experiment [13] [17]. For example, participants' motivation, attitude, character traits, beliefs, past experiences and the experimental environment (i.e. the setup of the laboratory and the experimental protocol) play an extremely important role in formation of emotion. Therefore, a comparison of emotional reactions induced by pictures and videos may be regarded as a hands-on tutorial for researchers as to which multimedia stimuli properties can lead to more accurate, precise and faster elicitation.

The study showed that standardized picture and sound databases can be joined together and used as videos for elicitation of high-arousal emotional states. Deliberate and practical stimulation of discrete emotions happiness and fear is attainable but it is more difficult and prone to error, especially happiness. The least successful was stimulation of very positive, i.e. high valence, emotional states.

We hope that the presented study could be used in design of emotion elicitation protocols as well as future affective multimedia databases. Additionally, the study's results may help researchers to find the optimal multimedia format for elicitation of emotion even if appropriate video stimuli are not available.